\documentclass{osa-article}
\graphicspath{{figs/}}

\journal{oe}

\begin{document}

\title{High-precision passive stabilization of repetition rate for a mode-locked fiber laser based on optical pulse injection}

\author{Tingting Yu,\authormark{1} Jianan Fang,\authormark{1} Qiang Hao,\authormark{2} Kangwen Yang,\authormark{2}  Ming Yan,\authormark{1,3} Kun Huang,\authormark{1,3,*} and Heping Zeng\authormark{1,3,4,5,$\dagger$}}

\address{\authormark{1}State Key Laboratory of Precision Spectroscopy, East China Normal University, Shanghai 200062, China\\
\authormark{2}School of Optical Electrical and Computer Engineering, University of Shanghai for Science and Technology, Shanghai 200093, China\\
\authormark{3}Chongqing Institute of East China Normal University, Chongqing 401121, China\\
\authormark{4}Jinan Institute of Quantum Technology, Jinan, Shandong 250101, China\\
\authormark{5}Shanghai Research Center for Quantum Sciences, Shanghai 201315, China}

\email{\authormark{*}khuang@lps.ecnu.edu.cn\\
\authormark{$\dagger$}hpzeng@phy.ecnu.edu.cn} 

\begin{abstract}
We have proposed and implemented a novel scheme to obtain high-precision repetition rate stabilization for a polarization-maintaining mode-locked fiber laser. The essential technique lies in the periodic injection of electronically modulated optical pulses into a nonlinear amplifying loop mirror within the laser resonator. Thanks to the nonlinear cross-phase modulation effect, the injected pulses referenced to an external clock serves as a stable and precise timing trigger for an effective intensity modulator. Consequently, synchronous mode-locking can be initiated to output ultrafast pulses with a passively stabilized repetition rate. The capture range of the locking system reaches to a record of 1 mm, which enables a long-term stable operation over 15 hours without the need of temperature stabilization and vibration isolation. Meanwhile, the achieved standard deviation is as low as 100 $\mu$Hz with a 1-s sample time, corresponding to a fluctuation instability of 5.0$\times10^{-12}$. Additionally, the repetition rate stabilization performance based on the passive synchronization has been systematically investigated by varying the average power, central wavelength and pulse duration of the optical injection.
\end{abstract}

\section{Introduction}
Mode-locked fiber lasers offer desirable features of compact size, long-time reliability, freeness of alignment, and flexible customization, which have thus been developed as an attractive light source in numerous fields including time-resolved spectroscopy, laser material processing, and nonlinear optical microscopy \cite{Kim2016AOP}. Especially, ultrafast fiber lasers with spectro-temporal engineering would facilitate broadened advanced functionalities requiring precise timing of pulse trains and accurate positioning of spectral modes \cite{Cundiff2003ROP}. For instance, the ultrashort pulse train with a regular interval is highly demanded in high-accuracy distance measurement, ultra-low noise microwave generation, high-precision timing synchronization, and high-capacity optical communication \cite{Kim2020JOSAA,Xie2017NP,Kim2010}. Moreover, simultaneous stabilization of the repetition rate and carrier envelope phase for a mode-locked laser would lead to realizing an optical frequency comb \cite{Diddams2020Science}, which provides a broadband phase-coherent optical source in an increasing number of metrological applications \cite{Newbury2011NP}. Therefore, it is important to develop robust techniques to precisely stabilize the repetition rate of a passively mode-locked fiber laser. 

In general, the locking of pulse repetition rate always relies on a certain way to compensate the optical-path variation of the laser cavity. The most common method is to control the geometric length, either mounting a movable mirror on a servo-controlled piezoelectric transducer (PZT) \cite{Zhang2009OE}, or attaching intra-cavity fibers on a PZT stretcher \cite{Washburn2004OL}. Typically, the available capture range of the active locking system is about tens of $\mu$m due to the limited ability of geometric elongation. Although the stretched amount could be extended by circling a long section of fiber around the piezoelectric actuator \cite{Walbaum2011APB}, yet inevitable bend-induced loss and birefringence might disturb the mode-locking operation. Additionally, the feedback bandwidth based on the PZT actuator is usually at the level of kHz, which may be insufficient to suppress high-frequency noises. Another practical limitation stems from the involved mechanical parts in these devices, which would degrade the long-term performance due to fiber abrasion or mechanical damage.

Alternatively, the stabilization of the round-trip optical path could be realized by controlling the refractive index of the propagation medium. This can be achieved via optical pumping of a dedicated doped fiber, where the resonantly-enhanced optical nonlinearity results in a significant refractive index change according to the Kramers-Kronig relation. In the seminal work, the all-optical approach was demonstrated to stabilize the repetition rate of an all-fiber Er-doped laser with a standard deviation of 22 mHz \cite{Rieger2013OE}. The residual fluctuation was then optimized to be 77 $\mu$Hz by driving the pump laser diode with a high-precision current source \cite{Hao2016JLT}. However, the capture range of the locking system was also limited to be tens of $\mu$m \cite{Yang2018LP}. Therefore, a long-term operation up to hours was obtained by either placing the active fiber on a thermo-electric element \cite{Rieger2013OE}, or containing the fiber laser in a temperature-controlled incubator \cite{Hao2016JLT}. Additionally, the feed-back bandwidth of the locking system is intrinsically restricted by the life-time of the excited state for the doped ion, which is typically at the ms level.

In parallel to the aforementioned active locking scheme, the passive fashion for controlling the intra-cavity pulse dynamics of ultrafast lasers has attracted increasing attention in recent years \cite{Chen2020COL}. In particular, the all-optical passive synchronization based on cross-phase modulation (XPM) has been implemented in various passive mode-locking configurations, \textit{e.g.}, based on nonlinear polarization rotation (NPR) \cite{Yoshitomi2006OL,Huang2012JSTQE,Tsai2013OL,Zhu2020AO}, saturable absorber (SA) \cite{Rusu2004OE,Zhang2011OL,Sotor2014OE,Li2020OL}, and nonlinear amplifying loop mirror (NALM) \cite{Huang2018OE,Zeng2019OL}. Furthermore, the optical synchronization to a master pulse source with a locked repetition rate was demonstrated to passively stabilize the slave laser cavity, where the stability and accuracy could be transferred in a simple but effective way \cite{Huang2018OE,Kuse2012OE}. In this case, the instantaneous nonlinear response enabled to access a high-bandwidth feedback, without the need of high-speed electronics and complicated feedback system in the active configuration. However, one remaining inconvenience is ascribed to the stringent requirement of careful cavity-length matching between two ultrafast lasers.

Here, we propose and implement an all-optical passive configuration to achieve a high-precision and robust repetition rate stabilization for a mode-locked fiber laser. In contrast to previous instantiations, an electrically modulated pulse source is used to provide a more flexible master injection at an arbitrary repetition rate. Counterintuitively, the long master pulses up to 1 ns is still possible for implementing a tight synchronization to an ultrashort fiber laser. The resultant standard deviation of repetition rate is about 100 $\mu$Hz for 1-s gate time, corresponding to a fluctuation instability of 5.0$\times10^{-12}$. Moreover, the capture range for the passive locking system is up to 1 mm, which is almost two orders of magnitude longer than that in conventional schemes. In combination with the all-polarization-maintaining fiber architecture, a long-term stable operation has been demonstrated over 15 hours without the need for temperature stabilization and vibration isolation.

\begin{figure}[t!]
\centering
\includegraphics[width=0.85\columnwidth]{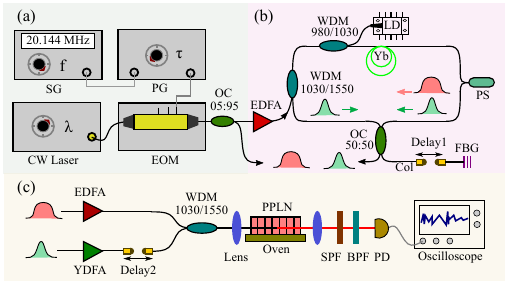}
\caption{Experimental scheme for all-optical and passive repetition rate stabilization of a mode-locked fiber laser. (a) Optical pulse generation based on the intensity modulation of a continuous-wave tunable laser at C-band wavelengths. (b) Injection locking of repetition rate for an Yb-doped mode-locked fiber laser. The injected master pulse within the nonlinear amplifying loop mirror propagates along with the clockwise slave pulse. The resultant non-reciprocal phase shift by the cross-phase modulation effect can initiate a stable synchronous mode-locking. Such passive and tight synchronization enables to stabilize the repetition rate with a high precision transferred from the signal generator (SG). (c) Schematic for the cross-correlation measurement of relative timing jitter.  PG: electrical pulse generator; EOM: electro-optic modulator;  OC: optical coupler; LD: Laser diode; WDM: wavelength division multiplexer;  Yb: ytterbium-doped gain fiber; FBG: fiber Bragg grating; PS: phase shifter;  Col: collimator; PPLN: periodically-poled lithium niobate crystal; SPF and BPF: short- and band-pass filer; PD: photodiode.}
\label{fig1}
\end{figure}

\section{Experimental setup}
Figure \ref{fig1} presents the experimental setup of the all-optical and passive stabilization system for locking the fiber-laser repetition rate. The whole system consisted of two parts: a master light source from electronically modulated pulses and a slave mode-locked Yb-doped fiber laser (YDFL), as shown in Figs. \ref{fig1}(a) and (b) respectively. The underlying mechanism for the repetition-rate stabilization lies in the light injection induced synchronous mode-locking. More specifically, The injected master pulses would serve as a stable and precise timing gate for initiating the formation of the mode-locked pulses in the salve laser cavity. Thanks to the XPM effect, the  cavity-length drift would be automatically compensated by the self-adapted spectral variation. Consequently, the round-trip optical path could be passively maintained due to the group velocity dispersion \cite{Furst1996JSTQE}. 

\begin{figure}[b!]
\centering
\includegraphics[width=0.8\columnwidth]{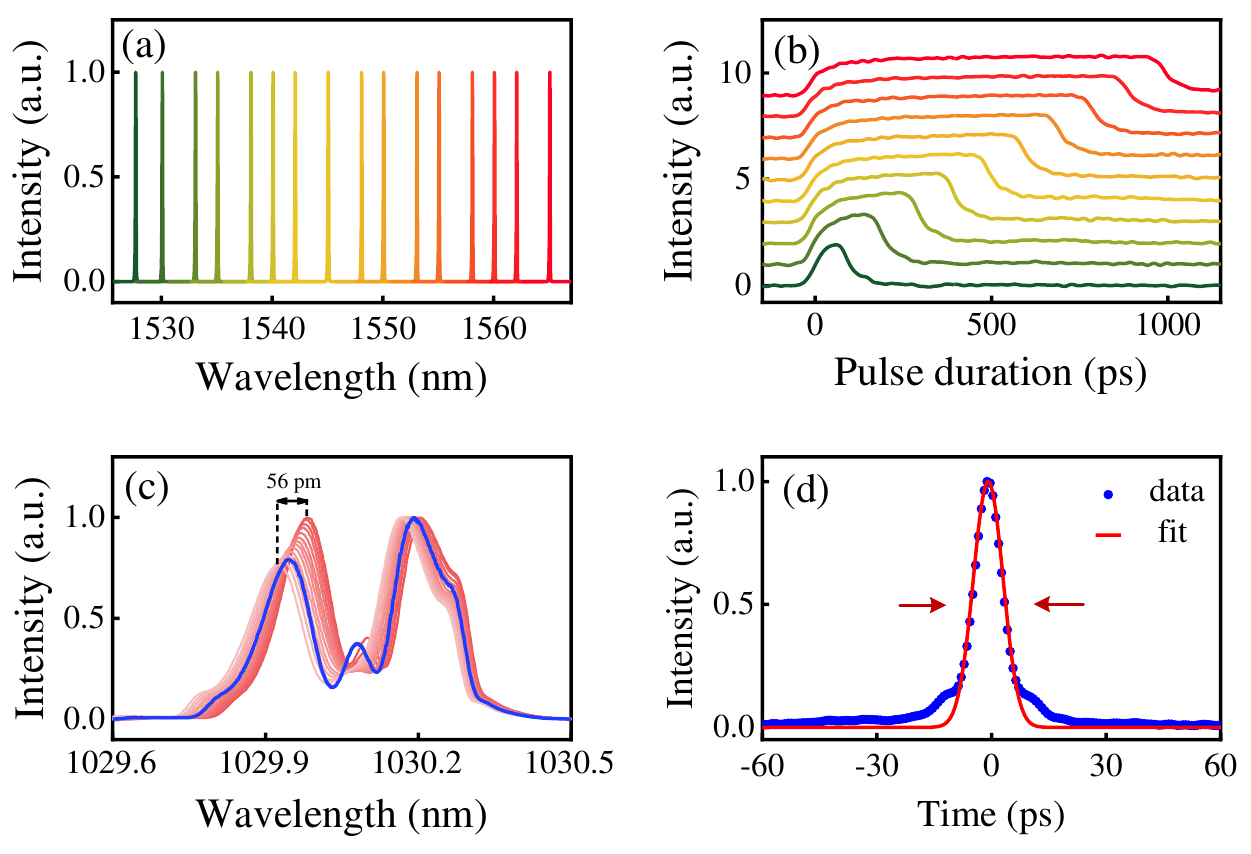}
\caption{Optical spectra (a) and temporal traces (b) for the electrically modulated pulse. The output wavelength could be tuned from 1528 to 1565 nm, while the pulse duration could be varied from 100 to 1000 ps. (c) Optical spectrum for the mode-locked optical pulse. Note that the curve in blue indicates the spectrum for the free-running operation, while the curves in red correspond to the spectrum for the synchronized state as varying the cavity length. (d) Autocorrelation trace for the laser pulse.}
\label{fig2}
\end{figure}

In our experiment, the master source originated from a continuous-wave (CW) wavelength tunable laser (LaseGen, LTL-1500). The output spectrum was in a narrow linewidth below 100 kHz, and the central wavelength could be tuned from 1527.6 to 1565.5 nm as shown in Fig. \ref{fig2}(a). Then, an electro-optic modulator (EOM) unit with a bandwidth of 10 GHz (Keyang Photonics, KY-MU-15-NRZ) was used to prepare the pulsed source. The EOM unit was equipped with an auto-bias controller to eliminate the long-term drift of the pulse intensity. The driving electrical pulse was provided by a tunable picosecond pulsar (LaseGen, LPP-200), which enabled to cover a wide pulse width tuning range from 50 to 1000 ps. The obtained optical pulse waveforms are illustrated in Fig. \ref{fig2}(b). Additionally, the repetition rate of electrically modulated pulses could be easily tuned by changing the output sinusoidal frequency in a signal generator (Rohde $\&$ Schwarz, SMC100A). The above flexibility of the master source in central wavelength, pulse duration, and repetition rate would facilitate the subsequent synchronization procedure, and open up broadened applications requiring a spectro-temporal tunability.

As for the slave laser shown in Fig. \ref{fig1}(b), the mode-locking mechanism was based on a nonlinear amplifying loop mirror (NALM). In the Sagnac loop, the clockwise and anticlockwise pulses interfere on the symmetric optical coupler (OC). The intensity-dependent nonlinear phase difference between the two propagating directions determines the transmission on the relevant pulse, which would act as a fast artificial saturable absorber \cite{Hao2016JLT}. In order to accumulate sufficient phase difference, the Yb-doped gain fiber was placed asymmetrically relative to the $2 \times 2$ balanced OC. Additionally, a $\pi$/2 non-reciprocal phase shifter was used to further reduce the mode-locking threshold \cite{Jiang2016PTL, Huang2018OE}. A fiber Bragg grating (FBG) with a bandwidth of 0.8 nm was connected to one output port of the OC, which could effectively provide an end mirror and a spectral filter. Under a pump power above 330 mW, self-starting mode-locked operation could be launched at a repetition rate about 20.144 MHz. The resulting spectrum centered at around 1030.2 nm with a bandwidth of 0.4 nm. The corresponding pulse duration was inferred to be about 6 ps from the Gaussian auto-correlation trace. 

To implement the passive synchronization, the repetition-rate difference of the master and slave pulses are required to be within the capture range. This requirement could be easily fulfilled by setting proper frequency value in the signal generator. Note that the two distance-adjustable collimators (Cols) denoted by delay1 in Fig. \ref{fig1}(b) was not necessary for matching the repetition rate. This delay line was used to characterize the synchronization system. To facilitate the optical synchronization, the master pulses were amplified to an average power up to 60 mW by an Er-doped fiber amplifier (EDFA). Finally, synchronous mode-locking could be initiated by simply injecting the amplified master pulses into the slave laser cavity through a wavelength division multiplexer (WDM). In this circumstance, only the clockwise slave pulse propagated along with the injected light as shown in Fig. \ref{fig1}(b), which resulted in an auxiliary non-reciprocal phase shift. Consequently, the optical injection would not only provide an effective timing gate for determining the origin of the mode-locked pulse formation, but also passively stabilize the pulse interval by the XPM-based pulling effect \cite{Wei2002APB}. 

Finally, we used a cross-correlation measurement for detailed investigation on the synchronization and stabilization performances. As shown in Fig. \ref{fig1}(c), the two-color pulses were spatially combined by a WDM before being focusing into a 25-mm periodically-poled lithium niobate (PPLN) crystal to perform the sum-frequency generation (SFG). The required quasi-phase matching for efficient nonlinear conversion was optimized at a poling period of 11.25 $\mu$m and an oven temperature of 34 $^\circ$C. The spectrally filtered SFG signal was then detected by a gain-variable photodiode (Thorlabs, PDA100A2).  A 1-MHz low-pass filter was applied on the output electronic signal such that the voltage amplitude was proportional to the SFG power. As a result, the fluctuation of the SFG signal would be linked to the relative timing jitter between the involved pulses. The related experimental results will be presented in the following section.

\begin{figure}[b!]
\centering
\includegraphics[width=0.5\columnwidth]{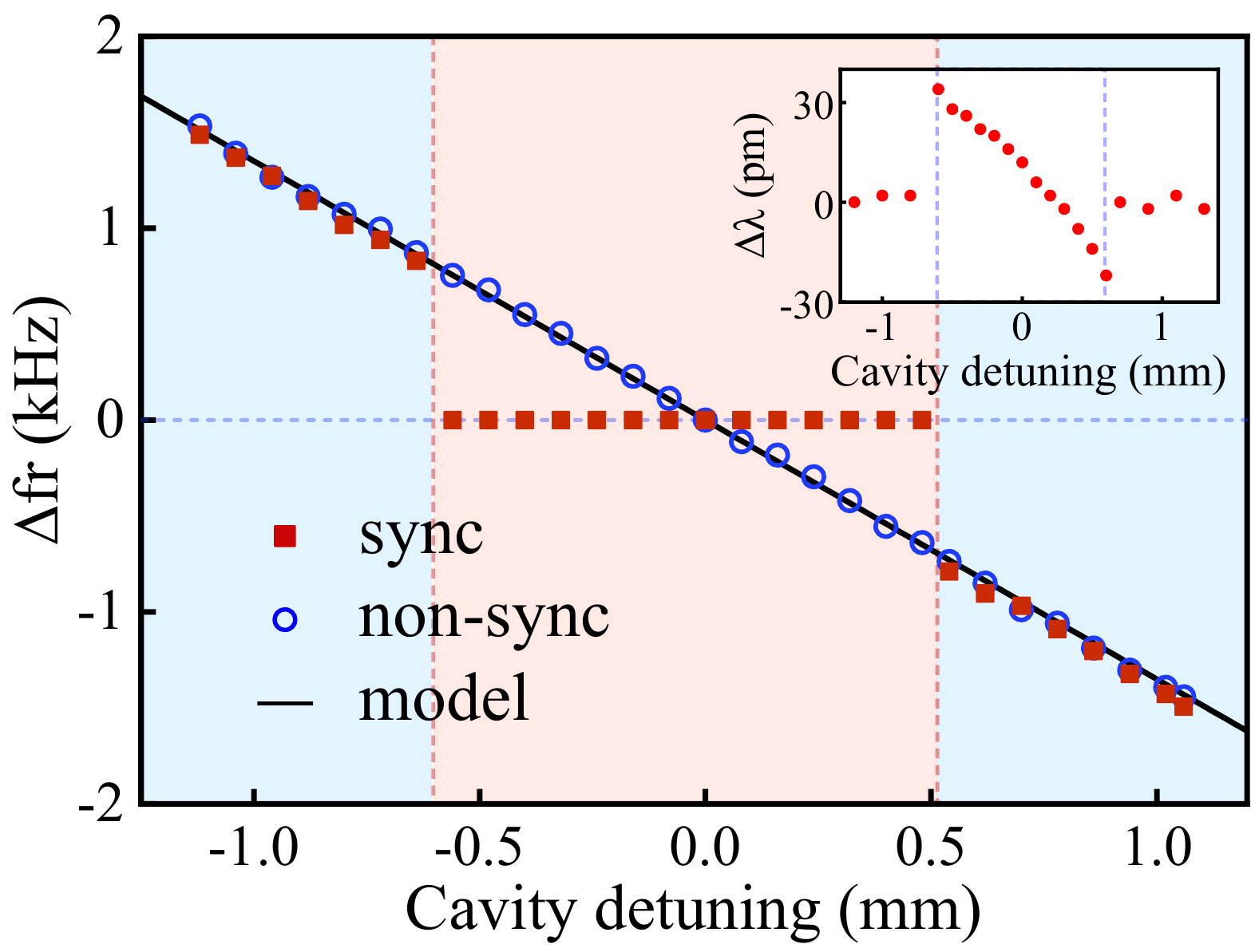}
\caption{Repetition-rate variation versus the cavity-length detuning of the fiber laser. Red solid squares and blue open circles correspond to synchronized and non-synchronized operation, respectively. The solid line is given by a theoretical model in the free-running regime. The inset shows the slight spectral shift as varying the cavity length, which is essential to achieve the passive synchronization based on group-velocity dispersion. }
\label{fig3}
\end{figure}

\section{Results and discussion}
Now we turn to characterize the implemented synchronization system. Figure \ref{fig3} presents repetition-rate changes $\Delta f_\text{r}$ as the slave cavity detuning $\Delta L$. In the absence of pulse injection, the repetition rate would decrease as elongating the cavity length according to the relation: 
\begin{equation}
\Delta f_\text{r} / f_\text{r0} = - \Delta L /L_0\ ,
\end{equation}
where $L_0$ denotes the round-trip optical path for the fiber laser at the repetition rate $f_\text{r0}$. With the optical injection, the repetition rate would jump to a constant value defined by the master pulses within the tolerance range of cavity-length mismatch. The resulting capture range of the passive stabilization system was found to be about 1 mm, which was almost two orders of magnitude longer than reported values obtained in previous active-locking configurations \cite{Rieger2013OE,Hao2016JLT,Yang2018LP}. The corresponding 1.5-kHz repetition-rate tolerance would be sufficient to accommodate the cavity-length drift for a fiber laser running in the open air, thus ensuring a long-term operation. The mechanism of the passive synchronization based on optical injection can be attributed to the XPM effect, which leads to instantaneous optical frequencies $\omega_1$ and $\omega_2$ for the dual-color pulses as given by \cite{Wei2002APB}
\begin{equation}
\begin{split}
\omega_1 &= \omega_{01} - n_2(\omega_{01}, \omega_{01}) \frac{\partial I_1}{\partial t} - n_2(\omega_{01}, \omega_{02}) \frac{\delta}{v_2} \frac{\partial I_2}{\partial t} \\
\omega_2 &= \omega_{02} - n_2(\omega_{02}, \omega_{02}) \frac{\partial I_2}{\partial t} - n_2(\omega_{02}, \omega_{01}) \frac{\delta}{v_1} \frac{\partial I_1}{\partial t} \ .
\end{split}
\end{equation}
Here $I_{1,2}$ is the pulse intensity, $n_2$ denotes the nonlinear index for the self-phase or cross-phase modulation, $v_{1,2}$ represents the mode volume, and $\delta$ is the overlap volume between the two lasers. It can be seen that the sign of the instantaneous frequency change for the slave pulse is opposite when the interaction time occurs at the leading or trailing part of the master pulse. It is the induced spectral shift that compensates cavity-length variation according to the group velocity dispersion, as shown in the inset of Fig. \ref{fig3}. The pulling effect between two pulses would automatically lead to maximal overlap again after enough round trips, thus exhibiting a dynamics process of self-synchronization \cite{Furst1996JSTQE,Wei2002APB}.

\begin{figure}[t!]
\centering
\includegraphics[width=0.75\columnwidth]{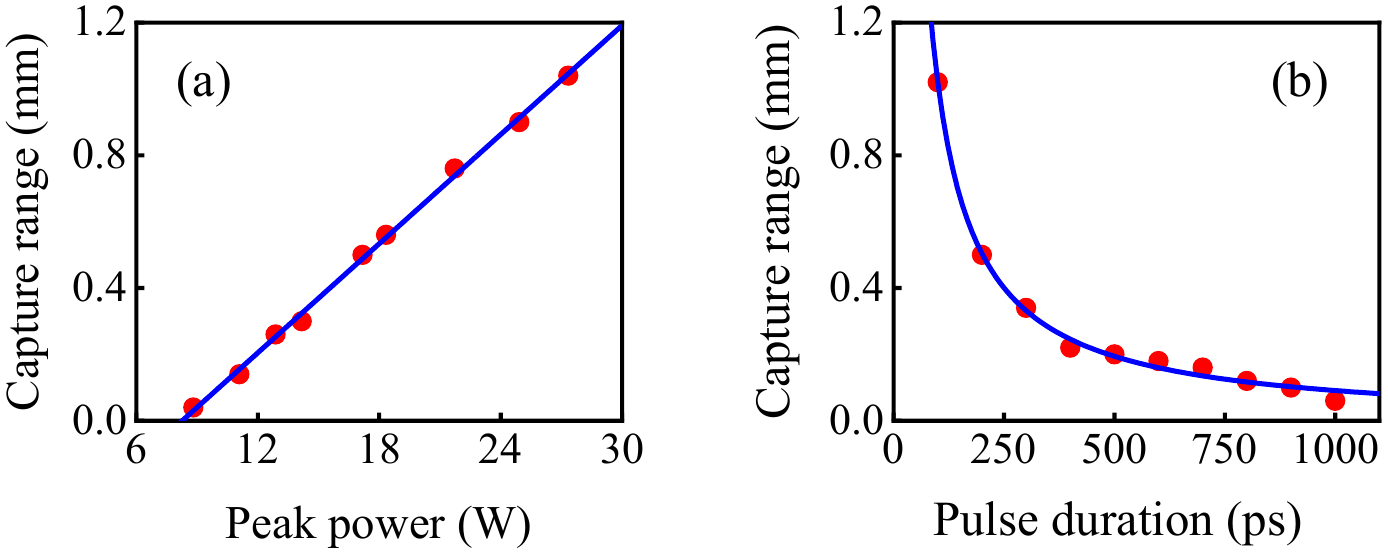}
\caption{(a) Capture range versus the peak power (a) and pulse duration (b) of the injection pulse. The solid lines are fitted according to a simplified model in the text.}
\label{fig4}
\end{figure}

Generally, the locking range of the pulse synchronization depends on the strength of nonlinear XPM effect, which would be related to the injected pulse energy.  As presented in Fig. \ref{fig4}(a), the cavity-length tolerance increased linearly as augmenting the pulse peak power. Specifically, the threshold of the peak power for launching a stable passive synchronization was found to be about 10 W, and the fitted slope was 55 $\mu$m/W. Taking into account the 1-mW output power of the mode-locked fiber laser, the minimum injection power of 18 mW for launching the synchronization operation would result in an injection ratio of 18. Additionally, the pulse width of the master injection was also investigated. Since the average power of the amplified pulses was kept almost constant in the experiment, the peak power as well as the corresponding capture range would be inversely proportional to the pulse duration. This expected behavior was illustrated in Fig. \ref{fig4}(b). 

\begin{figure}[b!]
\centering
\includegraphics[width=0.9\columnwidth]{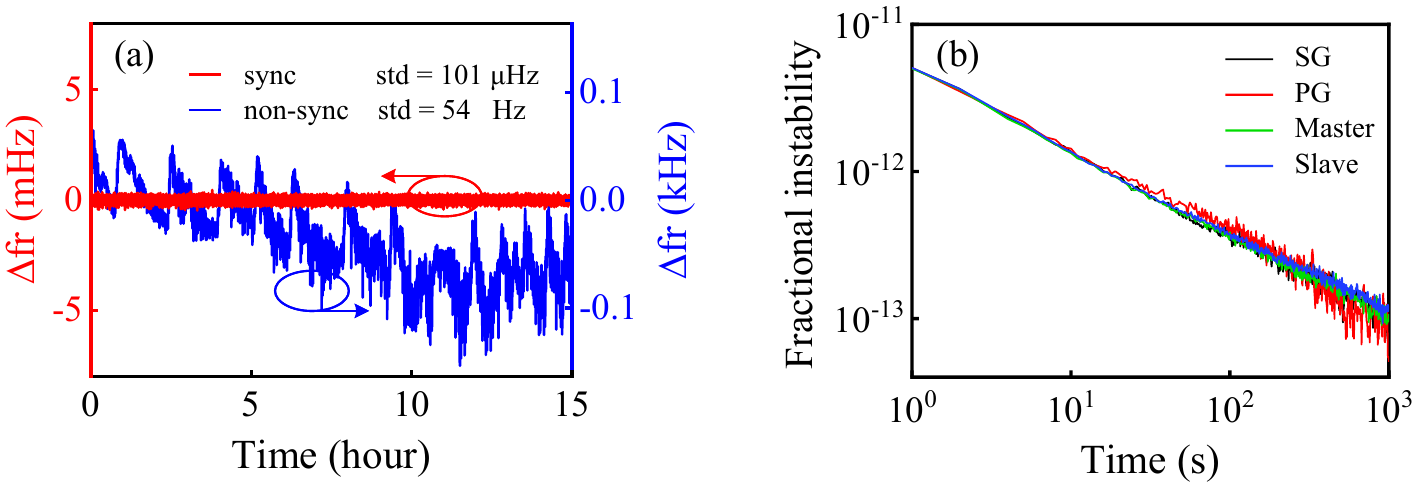}
\caption{(a) Repetition-rate change over 15 hours for the fiber laser at the synchronized and non-synchronized operation. The standard deviation (std) of the repetition rate is about 101 $\mu$Hz at the presence of injection locking. The sample time is set to be 1 s. (b) Calculated Allan deviation fractional instability of the repetition rate at different averaging time. The resultant behaviors are identical for the locked slave laser (blue), the master modulated pulse (green), the pulse generator (red), and the signal generator (black).}
\label{fig5}
\end{figure}

\begin{figure}[t!]
\centering
\includegraphics[width=0.65\columnwidth]{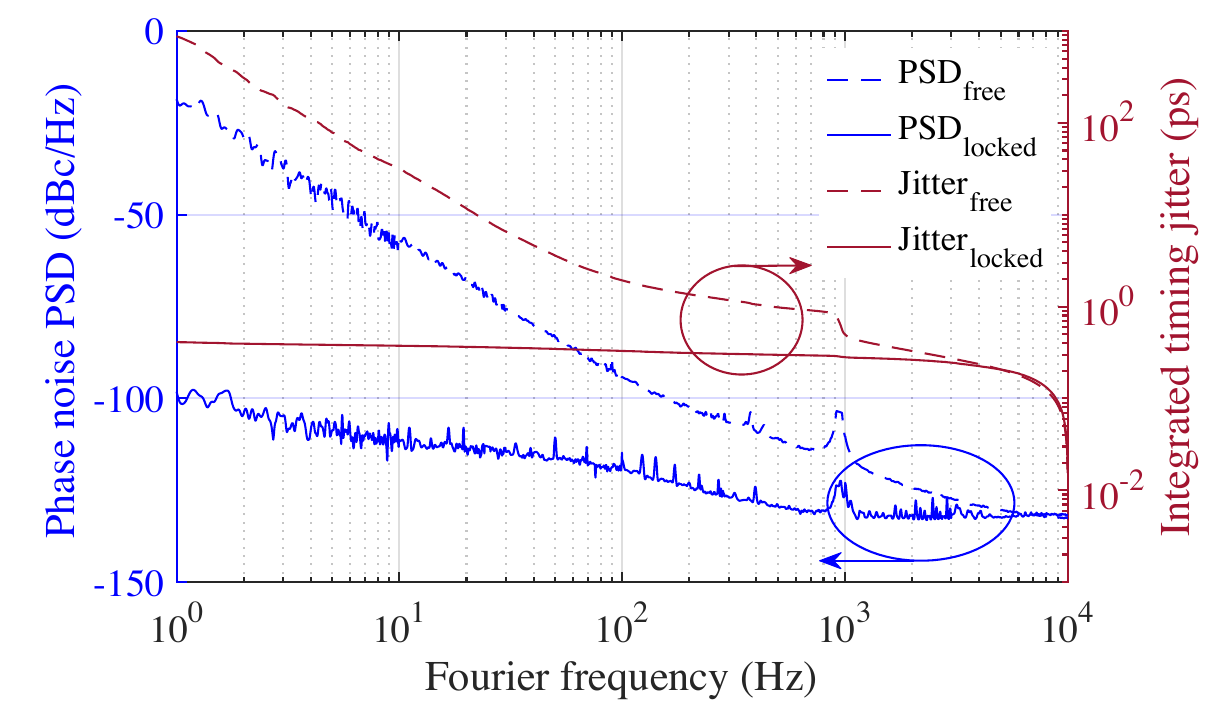}
\caption{Measured power spectral density of the phase noise for the fiber laser at the free-running and locked states. The corresponding integrated timing jitter is given within the Fourier frequency range from 10 kHz to 1 Hz.}
\label{fig6}
\end{figure}

In the following, we will further characterize the long-term performance of the passively stabilized fiber laser. For this purpose, the recorded pulse train by a 1-GHz photodiode was sent to a digital frequency counter (Tektronix FCA3103) with a sample time of 1 s. In all the involved measurement devices including the signal generator, frequency counter and digital oscilloscope, the clock was externally referenced to a hydrogen frequency standard (Microsemi, MHM2010). Figure \ref{fig5}(a) gives the result for a total acquisition time as long as 15 hours, which indicates a frequency fluctuation as small as 101 $\mu$Hz in terms of standard deviation. In contrast, the free-running fiber laser shows a large frequency drift about 200 Hz mainly due to the temperature variation in the lab. By considering a large capture range up to 1.5 kHz as shown in Fig. \ref{fig3}, the implemented synchronization system could in principle resist a much larger temperature variation. Assisted with the all-polarization-maintaining fiber structure, the presented system could potentially operate outdoors beyond the laboratory environment. To further quantify the repetition rate stability, the Allan deviation was calculated as a function of the average time $\tau$. As shown in Fig. \ref{fig5}(b), the behavior closed to $\tau^{-1/2}$ implies a dominant noise source from white frequency modulation \cite{Riley2008NIST}. In particular, the fractional instability at $\tau=1$ s was calculated to be 5.0$\times10^{-12}$, which was comparable to previous results \cite{Rieger2013OE, Hao2016JLT}. With a longer averaging time to 1000 s, the fractional instability would approach to 1.2$\times10^{-13}$. It is also worthy noting that the Allan deviation trace for the locked slave laser nearly coincided with those for the signal generator, master modulated pulses, and pulse generator as given in Fig. \ref{fig5}(b), which indicated a high-precision frequency transfer from the master electrical modulation to the slave optical pulse. To go beyond the measurement precision, we can mix the high-order harmonic repetition-rate frequency with a proper referenced signal, and send the resulting difference frequency component into the counter \cite{Hao2016JLT}. 

Furthermore, the output pulse from the mode-locked fiber laser was measured by a frequency analyzer (Keysight, N9320B), which indicated a fundamental repetition rate of 20.144 MHz and a signal-to-noises ratio of 70 dB. Moreover, the temporally stabilized pulse was characterized by a phase noise analyzer (Rohde \& Schwarz, FSWP). As shown in Fig. \ref{fig6}, the power spectral density (PSD) of the phase noise was effectively suppressed at the presence of the optical injection within the analysis spectral range between 1 Hz and 5 kHz. The integrated timing jitter within the Fourier frequency span from 1 Hz to 10 kHz was measured to be 0.41 ps.

\begin{figure}[t!]
\centering
\includegraphics[width=1\columnwidth]{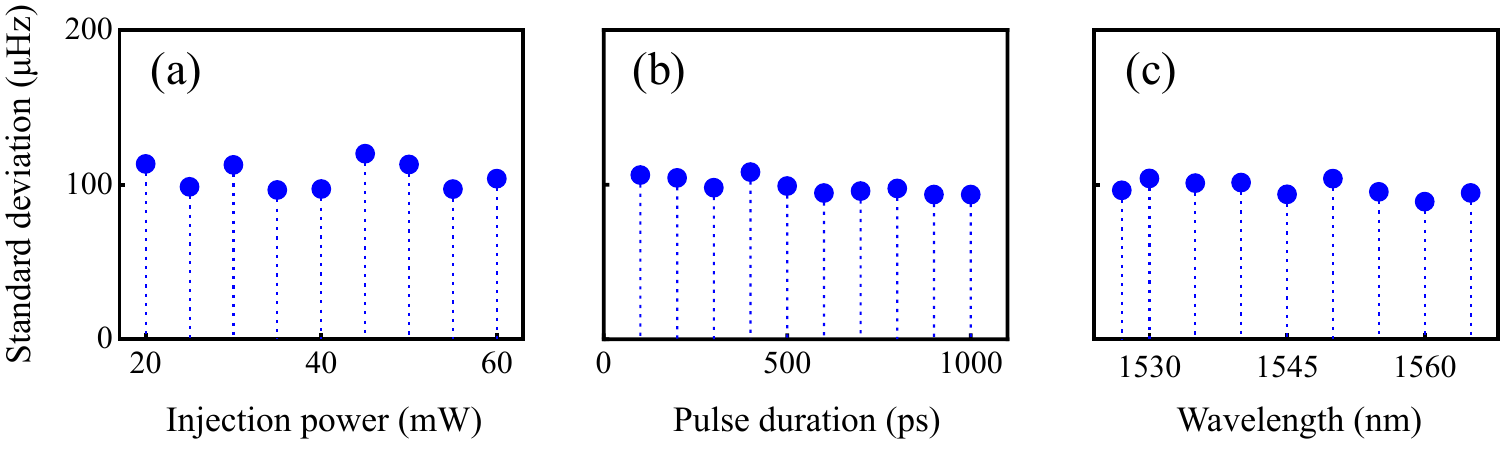}
\caption{Calculated standard deviation of the repetition rate for the locked fiber laser. Various parameters of the master pulse have been investigated by changing the average power (a), pulse duration (b), and central wavelength (c), respectively. Each data point is recored for 10 minutes with a 1-s sample gate.}
\label{fig7}
\end{figure}

\begin{figure}[b!]
\centering
\includegraphics[width=0.95\columnwidth]{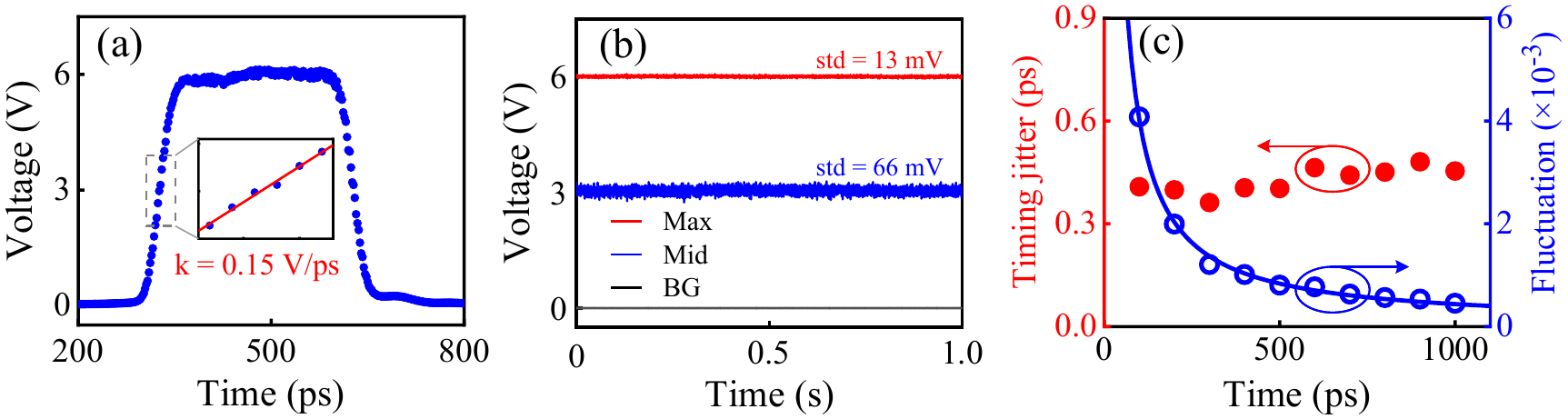}
\caption{(a) Cross-correlation trace between the two synchronized pulses. The linear part along the rising edge is zoomed in the inset. (b) Voltage traces are recorded at the maximum and middle points at the correlation trace. The background (BG) trace was measured by blocking the incident light. Each trace is acquired for 1 s at a sample rate of 1 MSa/s. (c) Calculated timing jitter for various injection pulse duration. The relative fluctuation is defined as the ratio between the timing jitter and master pulse duration.}
\label{fig8}
\end{figure}

Next, we have investigated the dependence of the repetition rate stability on pulse parameters for the master injection. As shown in Fig. \ref{fig7}, the achieved standard deviation could be maintained around 100 $\mu$Hz by changing the average power from 20 to 60 mW, pulse duration form 100 to 1000 ps, or central wavelength from 1528 to 1565 nm. Each data point was recorded for 10 minutes with a 1-s sample interval. Although the capture range for the locking system was reduced at the presence of lower injection peak power (\textit{i.e.,} smaller average power or longer pulse duration), the resulting stability of passive synchronization showed no sign of degradation. Further examination on the timing jitter between the synchronized pulses has also been conducted based on the optical cross-correlation technique \cite{Zeng2019OL}. Fig. \ref{fig8}(a) presents the obtained cross-correlation trace for a 200-ps master pulse as varying the time delay shown in Fig. \ref{fig1}(c). Near the half-maximum position, the temporal variation would be linearly translated into the intensity fluctuation. The temporal evolution of the SFG signal at the middle point was given in Fig. \ref{fig8}(b).The detector noise and intrinsic intensity fluctuation could be deduced from the voltage trace at the maximum. After correcting for this two contributions, the resulting the 53-mV variation would correspond to a timing jitter of 0.35 ps, which was much smaller than the cross-correlation width. Notably, the measured timing jitter could be maintained even for a 1000-ps master pulse. Consequently, the fluctuation relative to the pulse duration would reduced to 4.5$\times 10^{-4}$ as shown in Fig. \ref{fig8}(c).

\begin{figure}[t!]
\centering
\includegraphics[width=0.9\columnwidth]{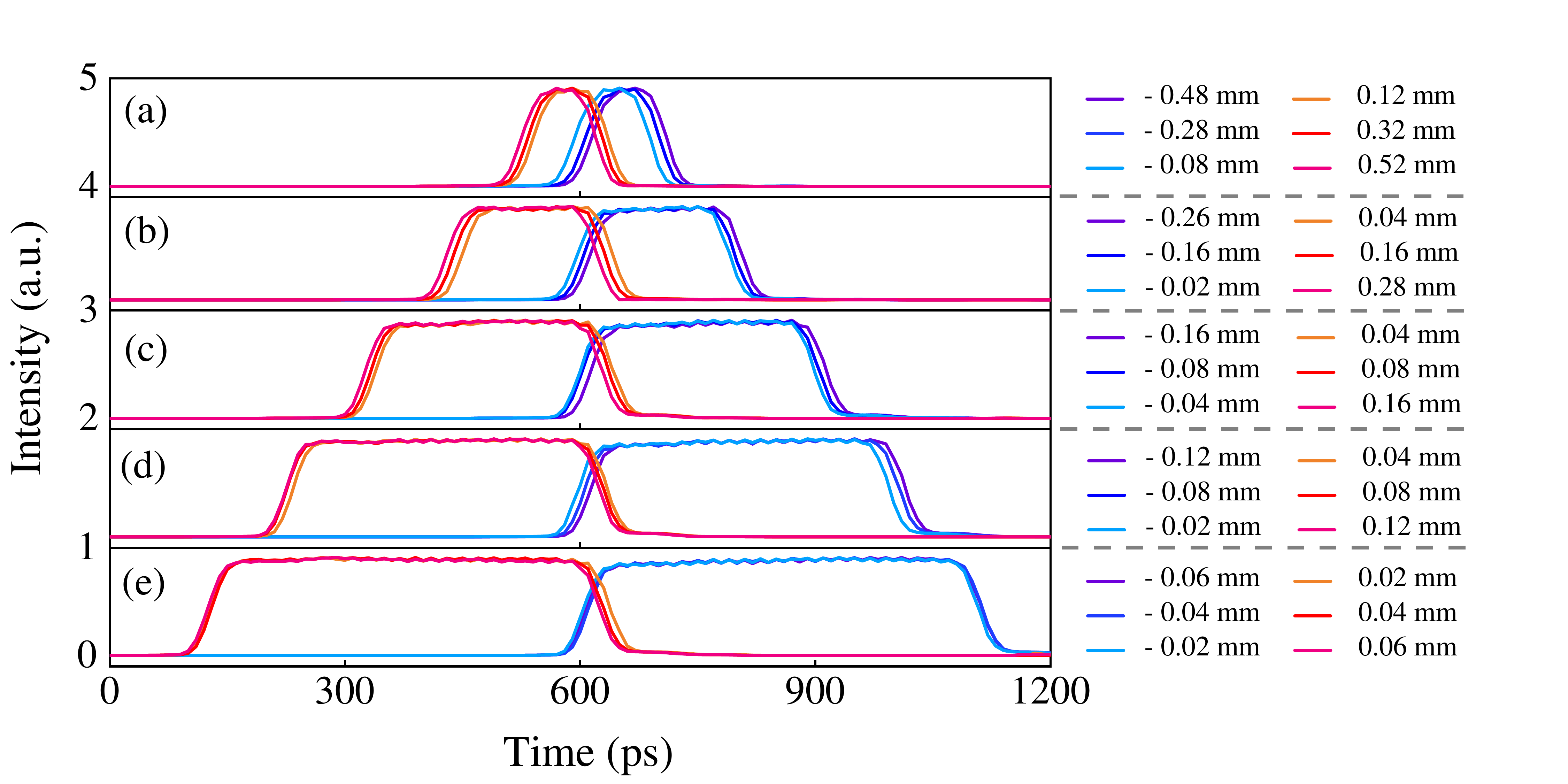}
\caption{Recorded cross-correlation trace as a function of delay time for various cavity-length detuning of the slave fiber laser. (a-e) show the similar behavior for various injection pulse duration from 100 to 500 ps. The temporal flip of the traces indicates the alternation between two distinct synchronous mode-locking states.}
\label{fig9}
\end{figure}

Finally, we experimentally confirmed that under the long pulse gating, the formation of the mode-locked pulse in the salve cavity tends to start at the leading or the trailing parts of the master pulse. Figure \ref{fig9} presents the recorded cross-correlation trace at different cavity-length detuning for the slave fiber laser. The amount of the temporal shift between the positive and negative detuning was significantly larger than that possibly induced by the length variation. This phenomenon became more prominent with injecting a long master pulse of 500 ps as shown in Fig. \ref{fig9}(e). Moreover, the temporal flip of the traces was found be close to the pulse width of the injection light, which indicated an alternation between two distinct synchronous mode-locking states originated at the leading or trailing edges of the master pulse. This intriguing behavior might be related to the fact that the pulse formation is more likely initiated at the presence of a dramatical transition between gain and loss.

\section{Conclusions}
To conclude, we have presented an all-optical passive configuration to lock the repetition rate of an Yb-doped mode-locked fiber laser. The superior stabilization performance was made possible by combining the robust optical timing synchronization and the stable electrically modulated pulses. In particular, the optical injection would not only facilitate the passive optical stabilization of the pulse interval based on the XPM-based pulling effect, but also provide an effective timing gate for determining the origin of the mode-locked pulse formation. Consequently, the capture range of the locking system reached to 1 mm, almost two orders of magnitude longer than the optical-path elongation accessed by the active techniques. Thanks to the full polarization-maintaining fiber architecture, the presented system exhibited a long-term operation over 15 hours. The obtained standard deviation for 1-s gate time was about 100 $\mu$Hz, corresponding to a fluctuation instability of 5.0$\times 10^{-12}$. The presented stabilization scheme may provide a simple and cost-effective way to implement a timing synchronization network \cite{Foreman2007}, where the master pulses could be easily distributed to various remote nodes. Moreover, the demonstrated setup is readily modified to other types of fiber lasers with erbium-, holmium- and thulium-doped gain medium. The involved optical gating would avoid the issue of lacking high-bandwidth integrated modulators in certain spectral regions, especially for actively mode-locked fiber lasers \cite{Wang2012OE}. Additionally, the master source was featured with a broadband tuning range in the central wavelength and pulse width, which might facilitate nonlinear frequency generation \cite{Murray2016OL,Huang2021PR} and Raman scattering spectroscopy \cite{Kong2020LSA}.

\section*{Funding.} 
National Key Research and Development Program (2018YFB0407100), Science and Technology Innovation Program of Basic Science Foundation of Shanghai (18JC1412000), Program for Professor of Special Appointment (Eastern Scholar) at Shanghai Institutions of Higher Learning, National Natural Science Foundation of China (11621404, 11727812), Shanghai Municipal Science and Technology Major Project (2019SHZDZX01), and Fundamental Research Funds for the Central Universities.

\section*{Acknowledgments.} 
The authors thank the insightful discussion with Dr. Y. Liu on the phase noise analysis. 

\section*{Disclosures.} 
The authors declare no conflicts of interest.

\section*{Data availability.} 
Data underlying the results presented in this paper are available from the corresponding author upon reasonable request.


\end{document}